\documentclass[twocolumn,showpacs,pra,superscriptaddress]{revtex4}

\usepackage{graphicx}

\newcommand{\Btop}{B_t}
\newcommand{\Bbias}{B_o}
\newcommand{\Bgrad}{B'}
\newcommand{\Bcurv}{B''}
\newcommand{\zo}{z_o}
\newcommand{\zpm}{z_\pm}
\newcommand{\Bb}{B_b}
\newcommand{\ro}{r_o}
\newcommand{\BH}{\Delta B}
\newcommand{\etal}{{\it et~al.}}

\begin{document}

\title{Double-well magnetic trap for {B}ose-{E}instein condensates}

\author{N. R. Thomas}
\affiliation{Physics Department, University of Otago, P.O. Box 56,
  Dunedin, New Zealand}

\author{C. J. Foot}
\affiliation{Clarendon Laboratory, Department of Physics, University of Oxford, \\
 Parks Road, Oxford, OX1 3PU, United Kingdom.}

\author{A. C. Wilson}
\affiliation{Physics Department, University of Otago, P.O. Box 56,
  Dunedin, New Zealand}
\date{August 10, 2001}

\begin{abstract}
  We present a magnetic trapping scheme for neutral atoms
  based on a hybrid of Ioffe-Pritchard and Time-averaged Orbiting Potential
  traps. The resulting double-well magnetic potential has readily controllable
  barrier height and well separation. This offers a new tool for
  studying the behavior of Bose condensates in double-well
  potentials, including atom interferometry and Josephson tunneling.
  We formulate a description for the potential of this magnetic
  trap and discuss practical issues such as loading with atoms, evaporative
  cooling and manipulating the potential.
\end{abstract}

\pacs{03.75.Fi, 39.25.+k, 85.70.Ay} 

\maketitle

\label{intro}

Bose-Einstein condensates (BECs) of dilute alkali gases have been the
subject of a great deal of attention since they were first realized   
\cite{Anderson1995a,Davis1995b,Bradley1995a}. One area in which there
has been considerable theoretical interest is 
the behavior of Bose condensates confined in double-well
potentials. The process of splitting a condensate in two has been
investigated by raising a potential barrier in  the center of a
harmonic trap
\cite{Menotti2001a,Spekkens1999a,Kagan1996a,Javanainen1999b,Capuzzi1999a}.
This work has given insight into the dynamic behavior of condensates
and modification of collective excitations. Also extensively studied
is the area of Josephson junctions
\cite{Javanainen1986a,Dalfovo1996d,Milburn1997a,Zapata1998a,Smerzi1997a,Williams2001},
where the condensate mean field affects the tunneling between wells
and macroscopic quantum self-trapping is possible. Other researchers
have proposed experiments involving the relative phase of the two
condensates. For example, uniting the relative phase of two initially
independent Bose condensates \cite{Jaksch2001,Molmer2001},
investigating quantum and thermal fluctuations of the phase
\cite{Pitaevskii2001}, and detection of weak forces \cite{Corney1999}. 

To date the only experimental demonstration of a double-well potential
for Bose condensates has been a hybrid configuration involving a
harmonic magnetic  potential divided into two regions by a repulsive
optical field \cite{Andrews1997b}. This scheme was used to observe
interference fringes of two overlapping Bose condensates, in the first
demonstration of long range coherence. More recently, the same
configuration was used to measure the presence of vortices induced in
one of the wells \cite{Inouye2001}. One feature of the apparatus is
that there is high degree of stability between the magnetic and
optical fields. 

In this work we address the problem of producing a purely magnetic
double-well potential suitable for studying BECs. Our scheme is an
extension of the existing Ioffe-Pritchard (IP) and Time-averaged
Orbiting Potential (TOP) traps used in many BEC experiments. We
begin by briefly summarizing the IP trap and review how a double
quadrupole-like potential can be formed. Bose condensates cannot be
confined in this potential, because of Majorana spin-flip loss, but we
show how the addition of a rotating bias field can resolve this
problem. The focus of this work is the development of a theoretical
description of the resulting double-well magnetic potential. Practical
issues such as loading, evaporative cooling and controlling the shape
of the potential are considered in detail.  

\section{Ioffe-Pritchard Traps}
\label{sec:IP}

\begin{figure}[bp]
  \begin{center}
    \includegraphics{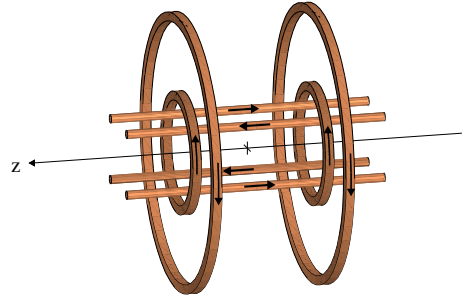}
    \caption{The layout of coils and bars for a Ioffe-Pritchard
      magnetic trap with current direction indicated by arrows. The
      four bars lie on corners of a square. Current in each
      bar flows in the direction opposite to that of its closest
      neighbors, and the magnitude of the current is the same for each
      bar. The two pairs of coils with large and small radii are the
      nulling and pinch coils respectively. The currents flow in the
      opposite sense so that they produce opposing magnetic fields.}
    \label{fig:IP_config}
  \end{center}
\end{figure}

Ioffe-Pritchard (IP) traps \cite{Pritchard1983a,Bergeman1987a}  have
been used extensively in the realization of BEC in alkali gases. The
basic configuration for an IP trap is illustrated in
Fig.~\ref{fig:IP_config}.
Four long ``bars'' with
currents in alternate directions  run parallel to the $z$ axis. These
generate a quadrupole field in the $x$-$y$ plane with gradient $\Bgrad$, which
leads to radial confinement. Two ``pinch'' coils have currents in
the same direction and are spaced to give a harmonic local minimum of the axial
magnetic field with curvature $\Bcurv$. These coils provide axial
confinement but also a large bias field $B_p$. Finally there are the
two ``nulling'' coils (usually in the Helmholtz configuration)
which provide a uniform axial field $B_n$. These allow the total bias
field at the trap center,
\begin{equation}
  \label{eq:Bbias}
  \Bbias = B_p - B_n,
\end{equation}
to be reduced.  For BEC experiments this is important because it leads
to tighter radial confinement and simplifies RF evaporation
\cite{Boyer2000a}. 

The magnetic field components for this geometry, to second order about
the trap center, are given by the equations  \hspace{1cm}
\begin{equation}
  \label{eq:field_components}  
  \begin{array}{lllll}
    B_x & = && +\Bgrad x  & - \frac{1}{2}\Bcurv x z,  \\
    B_y & = && -\Bgrad y  & - \frac{1}{2}\Bcurv y z, \\
    B_z & = &\Bbias  &&  + \frac{1}{2}\Bcurv (z^2 - \frac{1}{2}r^2),
  \end{array}
\end{equation}
where $r$ is the radial coordinate ($r^2 = x^2 + y^2$). The
terms with coefficient $\Bgrad$ are generated by the four bars.
Atoms in a weak-field seeking state are trapped at
the minimum of the field magnitude. For small thermal atomic clouds and
Bose-Einstein condensates ($k_B T < \mu \Bbias $), the
field magnitude is calculated by a binomial expansion to second order,
to give 
\begin{equation}
  \label{eq:B_IP}
  B_{\rm IP} = \Bbias + \frac{1}{2} \left( \frac{\Bgrad^2}{\Bbias} - \frac{\Bcurv}{2} \right) r^2  
  + \frac{1}{2}\Bcurv z^2,
\end{equation}
where $\Bbias > 0$.
The radial curvature is large when $ \Bgrad^2 / \Bbias
\gg \Bcurv/2 $, which is achieved by reducing $\Bbias$ with the
nulling coils. A summary of reported parameters for various
IP traps used in BEC experiments is shown in Table
\ref{tab:trap_param_table}. 
\begin{table}[tb]
  \begin{ruledtabular}
    \begin{tabular}{l|c|c|r|r|r}
      Ref.     & Variant & Species & $\Bbias$  & $\Bgrad$   & $\Bcurv$   \\
                    &         &         & (G)      &  (G/cm)    & (G/cm$^2$)       \\
      \hline
      \cite{Mewes1996a}             & Cloverleaf  & $^{23}$Na &  1   &  170 & 125 \\  
      \cite{Burt1997a}              & Baseball    & $^{87}$Rb &  1.6 &  300 & 85  \\  
      \cite{Sacket1997}*$\dagger$   & Perm. Magnet&    $^7$Li & 1000 & 1250& 855 \\  
      \cite{Hau1998a}*              & 4-Dee       & $^{23}$Na &  1.5 &  220 & 240 \\  
      \cite{Ernst1998a}             & Traditional & $^{87}$Rb &  1.5 &  275 & 365 \\  
      \cite{Esslinger1998a}         & QUIC        & $^{87}$Rb &  2   &  220 & 260 \\  
      \cite{Soding1999a}            & 3-coil      & $^{87}$Rb &  1.3 &  140 & 85  \\  
      \cite{Fried1998a}             & Superconductor &     H  &  0.5 &  240 &0.75 \\  
      \cite{Desruelle1999a}*        & Perm. Magnet& $^{87}$Rb & 190  &1200 & 400 \\  
      \cite{Arnold2001}*            & Baseball    & $^{87}$Rb &  1   &  175 & 62  \\  
      \cite{Torii2000a}             & Cloverleaf  & $^{87}$Rb &  1   &  175 & 185 \\  
      \cite{Fort2000a}*             & QUIC        & $^{87}$Rb &  1.6 &  170 & 110 \\  
      \cite{Aspect2001}*            & Cloverleaf  &$^4$He$^*$ &  0.3 &  85  &  40 \\  
      \cite{Leduc2001}              & QUIC        &$^4$He$^*$ &  4.2 &  280 & 200 \\  
    \end{tabular}
    \caption{A summary of reported IP trap parameters. In many cases the
      traditional geometry of Fig. \ref{fig:IP_config} is replaced by
      a variant with some specific advantages. References denoted *
      have field quantities calculated from the trap oscillation
      frequencies, and $\dagger$ indicates an equivalent radial
      gradient for traps that are harmonic.}
    \label{tab:trap_param_table}
  \end{ruledtabular}
\end{table}
Typical values for the popular coil-based IP traps are $\Bbias =
1$~G, $\Bgrad = 200$~G/cm and $\Bcurv =$~150 G/cm$^2$.  Note that the
curvature is usually much larger radially than axially so that atomic
clouds are {\em cigar} shaped with their long axis in the $z$
direction. For the values above, the radial curvature is $B_r''
\approx 40000 $ G/cm$^2$ and the anisotropy is $\lambda \equiv
\omega_r/\omega_z = \sqrt{B_r''/B_z''} = 16.3.$ In this calculation $\omega_z$ and
$\omega_r$ are the axial and radial oscillation frequencies in the harmonic trap.  

\begin{figure}[tb]
  \center
  \includegraphics[width=8.6cm]{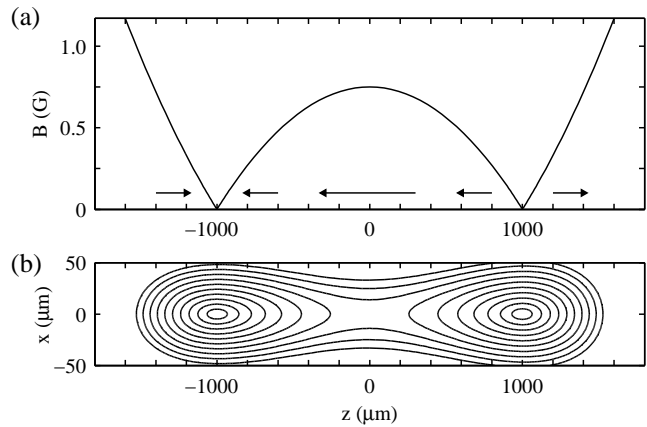}
  \caption{A double-well potential with well spacing of $2\zo = 2$~mm,
    generated by fields of $\Bbias = -$~0.75~G, $\Bgrad = $~200~G/cm
  and $\Bcurv =  $~150~G/cm$^2$. (a) The behavior
  for the $z$ axis with the field directions indicated by arrows. (b)
  Field magnitude contours in the $x$-$z$ plane with 0.1~G
  spacing. The barrier height is equivalent to a temperature (=$\mu_B
  \BH / k_B$) of 50~$\mu$K for atoms in a magnetic substate with $\mu = \mu_B$.} 
  \label{fig:double_well}
\end{figure}

Ketterle \etal~\cite{Ketterle1999a,QUICnote} have pointed out that a
double-well is formed if the nulling bias ($B_n$) is allowed to
overcome the pinch bias ($B_p$), so that $\Bbias < 0$ in
Eq.~(\ref{eq:Bbias}). The axial field now has zeros at the points $\zpm
= \pm \zo$, where 
\begin{equation}
  \label{eq:zo}
  \zo = \sqrt{\frac{2|\Bbias|}{\Bcurv}}.
\end{equation}
The trapping potential then has the form of two wells, as shown in
Fig. \ref{fig:double_well}.
The barrier height $\BH$ is $|\Bbias|$, which occurs at $(z,r)=(0,0)$. 

The field components, from Eq.~(\ref{eq:field_components}) evaluated
about the well bottom $\zpm$, are given by
\begin{eqnarray}
  \label{eq:field_components_overnulled}
  \begin{array}{llrlll}
    B_x & = &(   +\Bgrad - \frac{1}{2}\Bcurv \zpm)&x &&\\
    B_y & = &(   -\Bgrad - \frac{1}{2}\Bcurv \zpm)&y &&\\
    B_z & = &\Bcurv \zpm & z' & + & \frac{1}{2}\Bcurv  z'^2,
  \end{array}
\end{eqnarray}
where $z' = z - \zpm$. If we were to neglect the term in $z'^2$,
then these equations would be similar to those of a spherical
quadrupole field. 
This approximation is reasonable near the bottom of each well, where
$z' / 2\zpm \ll 1$. However, in general the curvature distorts the
potential from a quadrupole. Note that the axial gradient scales
with the well position $\zo$, so that confinement tightens and becomes
more linear as the well separation increases. The gradients along the
$x$ and $y$ axes are very similar because we will consider well
separations small enough that $\Bgrad \gg \Bcurv\zo$. The $x$-$y$
asymmetry is therefore small and can often be 
neglected, but we include it in subsequent calculations for
completeness. 

The usefulness of the potential in Fig.~\ref{fig:double_well} for BECs is
severely limited by Majorana spin flips at the two field zeros
\cite{Petrich1995a,Davis1995c}. Removing this means of trap loss in
the double-well trap above is the motivation behind this work.

\section{Double-TOP Trap}
\label{sec:double-top}

To eliminate loss from the two quadrupole-like wells, we apply the TOP trap scheme
developed by Cornell \etal~\cite{Petrich1995a} for the simple
quadrupole trap. We now consider adding a rotating bias field to
the double-well IP trap. The rotating field component, with magnitude
$\Btop$, is chosen to be  
\begin{equation}
  \label{eq:field_component_Btop}
  \begin{array}{lcc}
    B_x & = &  \Btop \, \cos{\omega t},\\
    B_y & = &  \Btop \, \sin{\omega t},\\
    B_z & = &  0.
  \end{array}
\end{equation}
The oscillating field has no $z$ component, which avoids the
introduction of further radial asymmetry and modulation
of the well spacing, stiffness, and barrier height during the bias
rotation. This choice also leads to a relatively simple description of
the trap. As in the standard TOP trap, the frequency of rotation
$\omega$ must satisfy 
\begin{equation}
  \omega_z,\omega_r \ll \omega \ll \omega_{L}
\end{equation}
where $\omega_{L}$ is the Lamor frequency in field $\Btop$. The
lower bound ensures a time-averaged potential, and the upper bound
allows the atomic magnetic dipole to follow the oscillating field
so that the atoms remain trapped. These limits cause no practical
difficulties as trap oscillation and Lamor frequencies are normally
of order 100 Hz and 1 MHz respectively.  

The key mechanism of the TOP trap is that the field zero of a
spherical quadrupole is displaced outside the atomic cloud by the
rotating bias field of Eq.~(\ref{eq:field_component_Btop}). In the
original implementation \cite{Petrich1995a} the field rotated about
the symmetry axis of a spherical quadrupole giving a circular locus
of $B=0$ at radius $\ro=\Btop/\Bgrad$, where $\Bgrad$ is the radial
gradient of the quadrupole. The double-TOP trap described in this work does
not in general have equal gradients in the plane in which the bias
rotates, so the locus is slightly elliptical. The axial displacements for $x$
and $y$ are given by 
\begin{equation}
  r_i = \frac{\Btop}{|B_i'|}
\end{equation}
where $i=\{x,y\}$ and $B_i' = \frac{\partial B_i}{\partial i}$ is the
magnetic field gradient on axis $i$ evaluated from
Eq.~(\ref{eq:field_components_overnulled}).
Some other implementations of the TOP trap 
\cite{Kozuma1999a,Muller2000a} also have an 
elliptical locus because the oscillating magnetic field does not
rotate about the symmetry axis of a spherical quadrupole.
The double-TOP will have a circular locus for the field zeros when the
$x$-$y$ asymmetry is negligible, corresponding to the limit $\zo \ll
1$~cm for our typical parameters.

\begin{figure}[tb]
  \begin{center}
    \includegraphics[width=8.6cm]{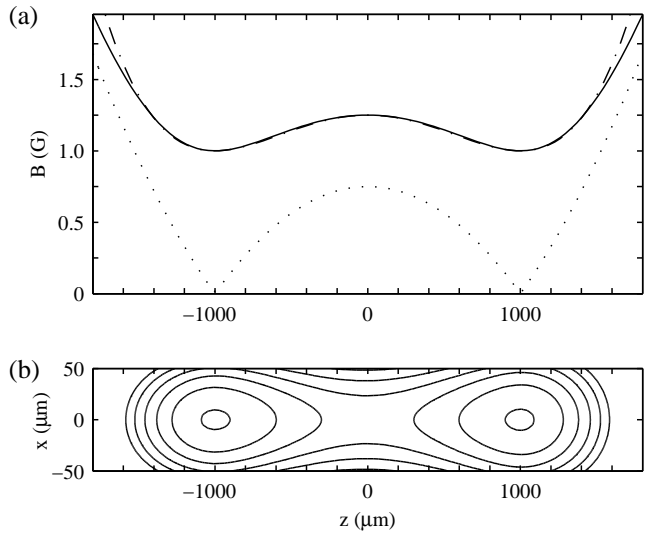}
    \caption{A double-TOP potential with the same parameters as
      Fig. \ref{fig:double_well} ($2\zo = 2$~mm , $\Bbias = -0.75$~G)
      and $\Btop = 1$~G. (a) The behavior on the $z$-axis with
      numerical (solid), Mexican-hat (dash-dot) and double-well IP
      (dotted) potentials shown. (b) A contour
      plot in the $x$-$z$ plane for the numerical integration
      with 0.1 G spacing starting from the well bottom $\Btop$. The
      barrier height is equivalent to 17 $\mu$K for atoms with $\mu = \mu_B$.}
    \label{fig:double_top}
  \end{center}
\end{figure}

Following Cornell's method for the standard TOP trap, the field
distribution for the double-TOP can be evaluated by averaging the
field magnitude: 
\begin{equation}
  \label{eq:average}
  B_{\rm av} = \frac{\omega}{2\pi}
  \int_{0}^{2\pi/\omega}{B(t) \ dt},
\end{equation}
where $B(t)$ is the instantaneous field magnitude. In
Fig.~\ref{fig:double_top} we have numerically evaluated
Eq.~\ref{eq:average}.
The addition of the rotating bias field to the double-well IP trap has
displaced the field zero as required, leaving a minimum of field magnitude of
$\Btop$ occurring at $(z,r)=(\zpm,0)$. The field at the top of the
barrier separating the two wells is now the quadrature sum of static and
rotating bias fields, given by
\begin{equation}
  \label{eq:Bb}
  \Bb = \sqrt{\Bbias^2 + \Btop^2}.
\end{equation}
The barrier height $\BH$ is then given by \mbox{$\BH = \Bb - \Btop$}. The
well separation is not affected by the addition of the rotating bias field.

There is no straightforward analytical solution to
Eq.~(\ref{eq:average}) that applies for a wide range of
parameters. However, it is possible to find results for restricted
conditions and we now present two such cases.  
Along the central axis of the trap, the field magnitude can be
approximated by a 1-dimensional Mexican-hat functional form  
\begin{equation}
  \label{eq:mex_hat}
  B_{\rm av}(z) = \Btop + \Delta B \left(1 - \left( \frac{z}{\zo} \right)^2 \right)^2,
\end{equation}
and the error is less than $1.5\%$  for $|\Bbias|/\Btop \leq
1$ in the region $ |z|/\zo \leq \sqrt{2}$. This description becomes
more accurate as the ratio $|\Bbias|/\Btop$ becomes
smaller. Figure~\ref{fig:double_top} shows that the fit is best in the
central region 
where $\Btop \leq B_{\rm av}(z) \leq \Bb$. At larger values of $z$ the
time-averaged field is dominated by the static axial component, so the
potential becomes harmonic. 

\begin{figure}[htb]
  \begin{center}
    \includegraphics[width=8.6cm]{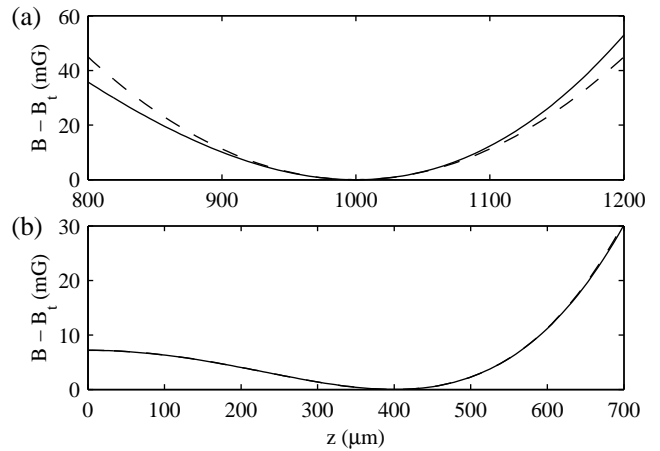}
    \caption{Right-hand well of double-TOP potentials with common field parameters
      $\Bgrad=$~200~G/cm, $\Bcurv=150$~G/cm$^2$ and $\Btop=1$~G. Both
      the numerical average (solid) and evaluation of
      Eq. \ref{eq:double_well_taylor_exp} (dashed) are shown. (a)
      The axial potential for widely separated wells
      ($\zo=$~1000~$\mu$m) and relatively high barrier
      ($\BH=$~250~mG). Eq.~\ref{eq:double_well_taylor_exp} is
      evaluated to second order.
      (b) The axial potential for closely separated wells
      ($\zo=$~400$\mu$m) and a low barrier ($\BH=$~7.2~mG). The
      analytical result is taken to 4th order in $z$.}
    \label{fig:double_top_approx}
  \end{center}
\end{figure}

The second case is to consider the shape of each well near the
minimum. This description applies for a cold atomic cloud confined
within either well. We proceed by performing an analytical
integration of Eq.~(\ref{eq:average}), using the double-well fields
expanded about $z=\pm \zo$ and the rotating bias. The resulting field
magnitude is given by
\begin{eqnarray}
  \label{eq:double_well_taylor_exp}
  B_{\rm av} \approx & \Btop \!\! & + \: \frac{1}{2}\!\!\left(\frac{\Bgrad^2}{2\Btop}
        +\frac{|\Bbias|\Bcurv}{4\Btop} \right)r^2 
        + \frac{\Bcurv\Bgrad}{4\Btop}\zpm (y^2-x^2) \nonumber \\
      & & + \frac{|\Bbias|}{\Btop}\Bcurv z'^2 
        + \frac{\Bcurv^2}{2\Btop}\zpm z'^3 + \frac{\Bcurv^2}{8\Btop}z'^4.
\end{eqnarray}
Examining this result we find that the radial
dependence is harmonic and is dominated by the term
$\Bgrad^2/(4\Btop)$, which is identical to the standard TOP trap with
field rotating in the $x$-$y$ plane.
The term in $(y^2-x^2)$ is the result of the slight $x$-$y$ asymmetry
of the double-well and is negligible for $\zo \ll 1$~cm.
Compared to the single-well IP trap of Eq.~(\ref{eq:B_IP}) the axial
curvature has been modified by the factor $2|\Bbias|/\Btop$, so that
axial confinement is tightened for $|\Bbias| \approx \Btop$. The terms in
$z'^3$ and $z'^4$ describe the barrier, and for our typical parameters
these terms are important when $\zo$ is smaller than approximately
500~$\mu$m. If the wells are more widely spaced, the third order term
adds only a slight tilt to the harmonic confinement and the 4th order term
can be neglected, as shown in Fig. \ref{fig:double_top_approx}.

In summary, the complete potential can be calculated numerically, while
near the well minima can be described analytically. The Mexican-hat
functional form is a useful approximation for the axial field
magnitude in the central region.

\section{Loading from a Ioffe-Pritchard trap}
\label{sec:loading}

The double-TOP trap may be loaded with a Bose condensate 
by transfer from a IP trap. To do this without spin-flip loss it is
necessary to apply the rotating bias before reverse biasing to form a
double-well. We model this loading scheme by adding the rotating bias
of Eq.~(\ref{eq:field_component_Btop}) to the IP field components of
Eq.~(\ref{eq:field_components}), and on integrating for the magnitude
find that
\begin{eqnarray}
  \label{eq:IP+RB}
  B_{\rm av} = & \Bb \;\; + & \frac{1}{2} \left[ \frac{\Bbias}{\Bb}    \left( \frac{\Bgrad^2}{\Bbias} - \frac{\Bcurv}{2}
  \right) - \frac{\Btop^2\Bgrad^2}{2\Bb^3} \right] r^2 \nonumber \\
      & & + \Bgrad \Bcurv \frac{\Bbias^2+\Bb^2}{4 \Bb^3} z(y^2-x^2)  \nonumber \\
      & & + \frac{1}{2}\frac{\Bbias}{\Bb}\Bcurv z^2 +  \frac{\Btop^2\Bcurv^2}{8\Bb^3}z^4.
\end{eqnarray}
Compared to the initial IP trap, both the axial and radial curvatures are
reduced by at least a factor of $\Bbias /\Bb$ due to the time-average. This
effect can be minimized if $\Bb \approx \Bbias$ which occurs when
$\Btop$ is small in comparison to $\Bbias$. Since $\Bbias > 0$, there
are no zeros of the field in this configuration and therefore there is no
restriction on the minimum size of $\Btop$. This choice of parameters
also minimizes the effect of the third term in the radial curvature. A
fourth order term in $z$ is required because the axial curvature goes to
zero with $\Bbias$.

\begin{figure}[htbp]
  \begin{center}
    \includegraphics{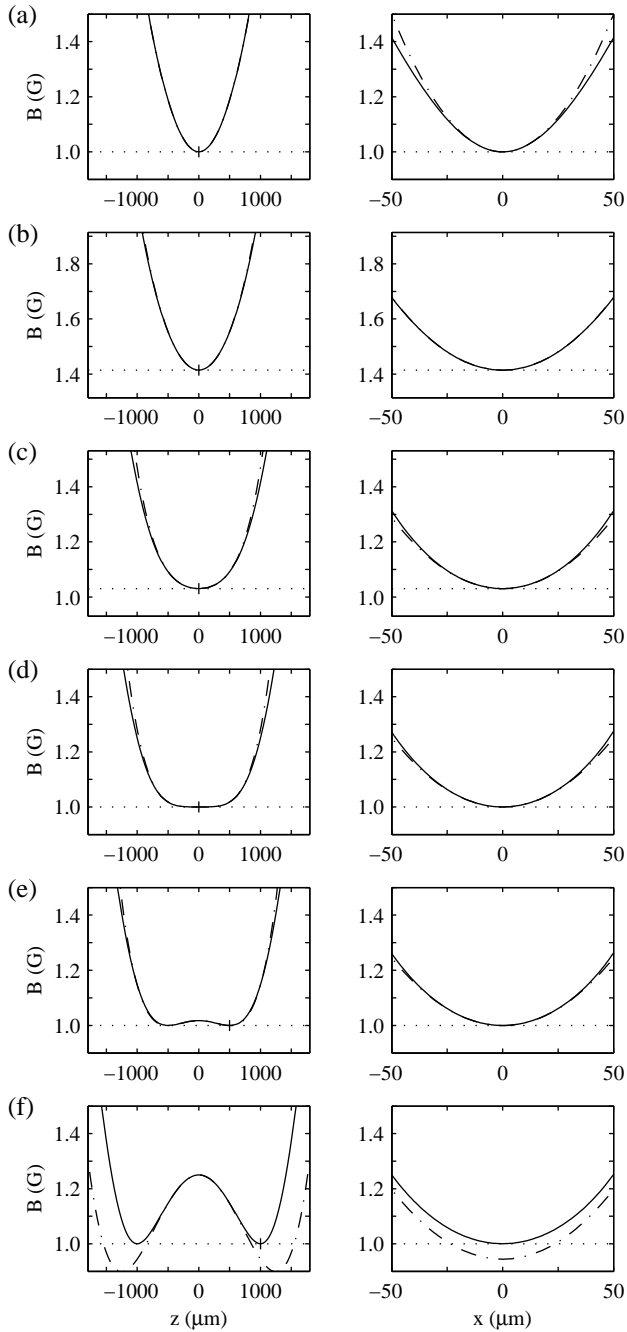}
    \caption{Numerical (solid) and analytical (dash-dot) modeling of
      the transformation from IP to double-TOP trap by reducing
      $\Bbias$. The behavior is shown for the
      $z$-axis (left column), and $x$-direction at the
      trap minima (right column)  indicated by the marker on the
      corresponding  $z$-axis plot. The plots show:
      (a) The initial IP trap with $\Bbias = 1$ G, $\Bgrad =$ 200
      G/cm, $\Bcurv =$ 150 G/cm$^2$, $\Btop = 0$~G; (b) The time-averaged IP
      formed with $\Btop = 1$~G; (c)-(f) $\Bbias$ equal to 0.25~G,
      0~G,$-0.19$~G, and $-0.75$~G respectively (all other parameters
      held constant).
      The well separations in (e) and (f) are $2\zo = 1000
      \mu$m and $2000 \mu$m. The radii of the locus traced by the two
      quadrupole field zeros in (d)-(f) are constant and equal to $r_x
      = 51.9$~$\mu$m and $r_y = 48.2$~$\mu$m.} 
    \label{fig:transform}
  \end{center}
\end{figure}

The loading process is an adiabatic evolution from IP to the
time-averaged IP trap and finally to double-TOP. We can use
Eq.~(\ref{eq:IP+RB}) to describe most of this process, as shown in
Fig. \ref{fig:transform}.
The first step is to increase $\Btop$ from 0 to 1~G while holding
$\Bbias = 1$~G. The trap minimum becomes $\Bb = \sqrt{2}\Bbias$, and
the axial and radial curvatures are reduced by factors of 0.71 and
0.53 respectively. A typical condensate is sufficiently small ($\approx
150$~$\mu$m long)
for the axial confinement to be harmonic. The second stage is to ramp
$\Bbias$ down to $-0.75$~G which creates a double-TOP trap characterized by
$\BH = 250$~mG, $2\zo =$~2000~$\mu$m. During this ramp the radial
confinement is mostly unchanged because the rotating bias dominates
$\Bb$. When $\Bbias$ passes through zero, $\Btop$ must be sufficiently
large to place the field zero outside the cloud of atoms and prevent
spin-flip loss. For the typical parameters used, the radial curvature
of $20000$ G/cm$^2$ at $\Bbias = 0$ means a typical alkali condensate
of $10^6$ atoms will be significantly inside $\ro = 50$ $\mu$m. The ramp of $\Bbias$ produces a
dramatic change in potential on the $z$ axis. While $\Bbias > 0$, the
curvature is positive but reducing, leading to a flattening as the
quartic term becomes more important. Some control of the rate of
change of the curvature is possible by the choice of $\Btop$ relative to
$\Bbias$. In this case, where they are approximately equal at start of
the ramp, the curvature does not greatly deviate from a linear
reduction, and this is also true for the case $\Btop > \Bbias$. In the
opposite extreme the curvature can collapse quite suddenly. When
$\Bbias$ passes through zero, the sign of the curvature reverses and
the barrier starts to rise. In  Fig. \ref{fig:transform}(f) we
see that although the analytical solution (Eq.~(\ref{eq:IP+RB}))
describes the qualitative shape of the potential, there is no longer
good quantitative agreement with the numerical result. The Mexican
hat description is more appropriate for this situation.

A second method for loading Bose condensates into a double-TOP
configuration is to evaporatively cool a thermal cloud. In this case
the evaporation process is much the same as for a standard TOP trap,
with the RF field oriented perpendicular to the rotating bias field,
except that RF resonance occurs at two points at the perimeter of the
cloud rather than one. 
Alternatively, an atomic cloud could be cooled first in a IP trap and
then in a double-TOP. However in this case, the standard
orientation for the RF antenna in a IP trap is not well matched for
optimum evaporation in the double-TOP.

\section{Controlling the barrier height and well spacing}
\label{sec:practical}

Careful control of the barrier height will be important in any
investigation of condensate splitting and Josephson tunneling. Such studies
are likely to require a barrier height comparable with the condensate chemical
potential and therefore much smaller than that which we considered in
the sections above. Recall that the barrier height is given by \mbox{$\BH =
\sqrt{\Bbias^2 + \Btop^2} - \Btop$}, so for the trap parameters associated with
Fig. \ref{fig:transform}(f) we have $\BH = 0.25$~G. For atoms trapped
in states with a magnetic dipole moment of $\mu_B$, this barrier
height is equivalent to a temperature of $\mu_B\BH/k_B = 16.8$~$\mu$K. In
comparison, the chemical potential, $\mu$ \cite{Dalfovo1999a}, of
10$^6$ atoms of $^{87}$Rb in the $|F=2,m_F=2\rangle$ state and confined
in one of the (almost harmonic) wells is $\mu/k_B =$~180~nK,
so that the ratio is $\mu_B\BH/\mu = 93$. 

\begin{figure}[tb]
  \begin{center}
    \includegraphics{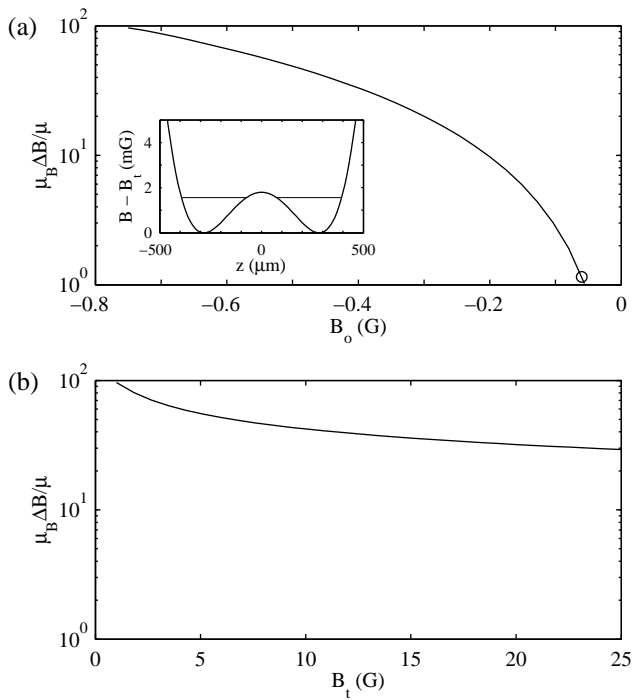}
    \caption{(a) Control of the barrier height by altering the static bias field
      $\Bbias$ with $\Btop = 1$ G. In the range shown, the ratio
      of barrier height and chemical potential drops from 100 to
      1. The circle indicates the parameters used for the
      inset. Inset: the magnetic potential for $\Bbias = -60$~mG and
      the chemical potential (horizontal lines) for 10$^6$ Rb atoms in
      each well. The calculation of $\mu$ is approximated by
      neglecting any effect of the condensate in 
      the other well.
      (b) Control of the barrier height by altering the rotating bias
      field $\Btop$ with $\Bbias = -0.75$~G. Note that even at 25~G
      the barrier height is still considerably greater than the
      chemical potential.} 
    \label{fig:barrier_height}
  \end{center}
\end{figure}

The options for lowering the
barrier height are to lower $\Bbias$ or increase $\Btop$. Increasing $\Btop$
keeps the well positions fixed but relaxes the radial confinement,
while decreasing $|\Bbias|$ reduces the well spacing. In
Fig.~\ref{fig:barrier_height}(a) we show the change in barrier height
as a function of $\Bbias$ while keeping $\Btop$ fixed. 
Initially the barrier is some 100 times higher than the chemical
potential but decreases to a comparable height at $\Bbias =
-55$~mG. The chemical potential is estimated by a numerical method
where we integrate over the magnetic potential in the Thomas-Fermi
approximation \cite{Dalfovo1999a}. We keep the number of atoms
constant at $10^6$ for each value of $\Bbias$ and solve for the
chemical potential. Note that achieving very low barriers with this
method requires very small values of $\Bbias$, which is discussed below.  

We could also lower the barrier height by increasing $\Btop$, as shown
in Fig.~\ref{fig:barrier_height}(b). For our choice of parameters,
using this method to make the barrier height equivalent to the
chemical potential would involve increasing $\Btop$ to
approximately $1 \times 10^5$~G, and is therefore
impractical. However, with a lower value of $\Bbias$, increasing
$\Btop$ offers greater sensitivity.  For example, with $ \Bbias =
-110$~mG the barrier height and chemical potential become comparable
when $\Btop$ is approximately 20~G.

Achieving small well spacing also requires a very small bias field.
For example, to obtain a well spacing of \mbox{$2\zo=300$ $\mu$m}
requires a static bias field of only -17~mG for our choice of axial
curvature. Note that \mbox{$\zo = \sqrt{2|\Bbias| / \Bcurv}$}, so that
best control is achieved for largest possible curvatures and widely
separated wells. 

Achieving tight control of close well spacing and low barrier height
therefore requires a high degree of coil current stability and
shielding from external fields. High current power supplies of the
sort often used for IP traps have relatively poor current
stability, so that achieving close well separation with these is 
likely to be difficult. However, a low current example where a high
degree of stability has been achieved is the QUIC trap
\cite{Esslinger1998a}. The authors report in Ref. \cite{Bloch1999a} reducing
``residual fluctuations in the magnetic field to a level below 0.1
mG''. Since the bias field required to achieve control over the well
separation is small it should be possible to add a small external bias
field without compromising the stability of a QUIC-style
trap. It is worth noting that in the limit of a very small well spacing it becomes
increasingly difficult to form a significant barrier, because this requires
$\Btop \ll \Bbias$. This condition leads to a very small $\ro$ and
subsequent trap loss from Majorana spin-flips. Ensher has also found
evidence \cite{Ensher1998a} that an atomic cloud in a TOP trap with
small $\Btop$ is vulnerable to transitions  to untrapped magnetic spin
states, induced by residual AC magnetic fields  associated with
current noise. 

\section{Conclusions}
\label{sec:conclusion}

We have developed a theoretical description for a double-well magnetic
trap suitable for confining Bose-Einstein condensates and cold thermal
clouds. The separation of the wells is controlled by the static bias field
$\Bbias$, while the barrier height also depends on the rotating bias
field $\Btop$. We have developed analytical forms of the potential
that describe one or both wells, which are in good agreement with
numerical simulations of the time-averaged field. 
This trap has the convenient feature that it is based solely on
magnetic fields and therefore avoids the issue of the relative stability
of magnetic and optical fields addressed in \cite{Andrews1997b}.  It also
offers a barrier height and well spacing that are tunable, so that a
wide range of condensate phenomena may be accessible. Wide well
separations are readily achieved and may be useful for interferometric
applications, where a condensate in one well is perturbed and a second
(local oscillator) is not. The double-TOP scheme also has the
advantage that it is based on existing trapping technologies.

\begin{acknowledgments}
We are acknowledge the support of the Marsden Fund, contract UOO910,
and the University of Otago.
\end{acknowledgments}


\begin{thebibliography}{46}

\bibitem{Anderson1995a}
M.~H. Anderson {\it et~al.}, Science {\bf 269},  198  (1995).

\bibitem{Davis1995b}
K.~B. Davis {\it et~al.}, Phys. Rev. Lett. {\bf 75},  3969  (1995).

\bibitem{Bradley1995a}
C.~C. Bradley, C.~A. Sackett, J.~J. Tollett, and R.~G. Hulet, Phys. Rev. Lett.
  {\bf 75},  1687  (1995); {\bf 79}, 1170 (1997).

\bibitem{Menotti2001a}
C. Menotti, J.~R. Anglin, J.~I. Cirac, and P. Zoller, Phys. Rev. A {\bf 63},
  023601  (2001).

\bibitem{Spekkens1999a}
R.~W. Spekkens and J.~E. Sipe, Phys. Rev. A {\bf 59},  3868  (1999).

\bibitem{Kagan1996a}
Y. Kagan, E.~L. Surkov, and G.~V. Shlyapnikov, Phys. Rev. A {\bf 54},  R1753
  (1996).

\bibitem{Javanainen1999b}
J. Javanainen and M.~Y. Ivanov, Phys. Rev. A {\bf 60},  2351  (1999).

\bibitem{Capuzzi1999a}
P. Capuzzi and E.~S. Hern{\'a}ndez, Phys. Rev. A {\bf 59},  3902  (1999).

\bibitem{Javanainen1986a}
J. Javanainen, Phys. Rev. Lett. {\bf 57},  3164  (1986).

\bibitem{Dalfovo1996d}
F. Dalfovo, L. Pitaevskii, and S. Stringari, Phys. Rev. A {\bf 54},  4213
  (1996).

\bibitem{Milburn1997a}
G.~J. Milburn, J. Corney, E.~M. Wright, and D.~F. Walls, Phys. Rev. A {\bf 55},
   4318  (1997).

\bibitem{Zapata1998a}
I. Zapata, F. Sols, and A.~J. Leggett, Phys. Rev. A {\bf 57},  R28  (1998).

\bibitem{Smerzi1997a}
A. Smerzi, S. Fantoni, S. Giovanazzi, and S.~R. Shenoy, Phys. Rev. Lett. {\bf
  79},  4950  (1997).

\bibitem{Williams2001}
J.~E. Williams, Phys. Rev. A {\bf 64},  013610  (2001).

\bibitem{Jaksch2001}
D. Jaksch {\it et~al.}, Phys. Rev. Lett. {\bf 86},  4733  (2001).

\bibitem{Molmer2001}
K. Molmer, e-print cond-mat/0105533.

\bibitem{Pitaevskii2001}
L. Pitaevskii and S. Stringari, e-print cond-mat/0104458.

\bibitem{Corney1999}
J.~F. Corney, G.~J. Milburn, and W. Zhang, Phys. Rev. A {\bf 59},  4630
  (1999).

\bibitem{Andrews1997b}
M.~R. Andrews {\it et~al.}, Science {\bf 275},  637  (1997).

\bibitem{Inouye2001}
S. Inouye {\it et~al.}, e-print cond-mat/0104444.

\bibitem{Pritchard1983a}
D.~E. Pritchard, Phys. Rev. Lett. {\bf 51},  1336  (1983).

\bibitem{Bergeman1987a}
T. Bergeman, G. Erez, and H.~J. Metcalf, Phys. Rev. A {\bf 35},  1535  (1987).

\bibitem{Boyer2000a}
V. Boyer {\it et~al.}, Phys. Rev. A {\bf 62},  021601(R)  (2000).

\bibitem{Mewes1996a}
M.-O. Mewes {\it et~al.}, Phys. Rev. Lett. {\bf 77},  416  (1996).

\bibitem{Burt1997a}
E.~A. Burt {\it et~al.}, Phys. Rev. Lett. {\bf 79},  337  (1997).

\bibitem{Sacket1997}
C.~A. Sackett, C.~C. Bradley, M. Welling, and R.~G. Hulet, App. Phys. B {\bf
  655},  433  (1997).

\bibitem{Hau1998a}
L.~V. Hau {\it et~al.}, Phys. Rev. A {\bf 58},  R54  (1998).

\bibitem{Ernst1998a}
U. Ernst {\it et~al.}, Europhys. Lett. {\bf 41},  1  (1998).

\bibitem{Esslinger1998a}
T. Esslinger, I. Bloch, and T.~W. H{\"a}nsch, Phys. Rev. A {\bf 58},  R2664
  (1998).

\bibitem{Soding1999a}
J. S{\"{o}}ding {\it et~al.}, Appl. Phys. B {\bf 69},  257  (1999).

\bibitem{Fried1998a}
D.~G. Fried {\it et~al.}, Phys. Rev. Lett. {\bf 81},  3811  (1998).

\bibitem{Desruelle1999a}
B. Desruelle {\it et~al.}, Phys. Rev. A {\bf 60},  R1759  (1999).

\bibitem{Arnold2001}
A.~S. Arnold, C. MacCormick, and M.~G. Boshier, e-print cond-mat/0103586.

\bibitem{Torii2000a}
Y. Torii {\it et~al.}, Phys. Rev. A {\bf 61},  041602(R)  (2000).

\bibitem{Fort2000a}
C. Fort {\it et~al.}, Europhys. Lett. {\bf 49},  8  (2000).

\bibitem{Aspect2001}
A. Robert {\it et~al.}, Science {\bf 292},  461  (2001).

\bibitem{Leduc2001}
F. {P}ereira~{D}os Santos {\it et~al.}, Phys. Rev. Lett. {\bf 86},  3459
  (2001).

\bibitem{Ketterle1999a}
W. Ketterle, D.~S. Durfee, and D.~M. Stamper-Kurn,  in {\em Proceedings of the
  International School of Physics - Enrico Fermi}, edited by M. Inguscio, S.
  Stringari, and C. Wieman (IOS Press, Amsterdam, 1999), p.\ 67.

\bibitem{QUICnote}
A very similar double-well potential is used routinely when loading
laser cooled atoms into a QUIC style Ioffe-Pritchard trap \cite{Esslinger1998a}.

\bibitem{Petrich1995a}
W. Petrich, M.~H. Anderson, J.~R. Ensher, and E.~A. Cornell, Phys. Rev. Lett.
  {\bf 74},  3352  (1995).

\bibitem{Davis1995c}
K.~B. Davis {\it et~al.}, Phys. Rev. Lett. {\bf 74},  5202  (1995).

\bibitem{Kozuma1999a}
M. Kozuma {\it et~al.}, Phys. Rev. Lett. {\bf 82},  871  (1999).

\bibitem{Muller2000a}
J.~H. M{\"u}ller {\it et~al.}, J. Phys. B {\bf 33},  4095  (2000).

\bibitem{Dalfovo1999a}
See, for example F. Dalfovo, S. Giorgini, L.~P. Pitaevskii, and S. Stringari, Rev. Mod. Phys.
  {\bf 71},  463  (1999).

\bibitem{Bloch1999a}
I. Bloch, T.~W. H{\"a}nsch, and T. Esslinger, Phys. Rev. Lett. {\bf 82},  3008
  (1999).

\bibitem{Ensher1998a}
J.~R. Ensher, Ph.D. thesis, University of Colorado (Boulder), 1998.

\end{thebibliography}
\end{document}